\documentclass[10pt]{article}
\usepackage{amsmath}
\usepackage{amssymb}
\usepackage{natbib}

\begin{document}

\bibliographystyle{plainnat}
\setcitestyle{numbers,square}

\title{\large{A numerical method for computing optimum radii of host stars \\
        and orbits of planets, with application to Kepler-11, Kepler-90, \\     
        Kepler-215, HD~10180, HD~34445, and TRAPPIST-1}}

\author{Vassilis S. Geroyannis \\
        Department of Physics, University of Patras, Greece          \\  
        vgeroyan@upatras.gr}
\date{September 10, 2020}

\maketitle

\begin{abstract}
In the so-called ``global polytropic model'', we assume planetary systems in hydrostatic equilibrium and solve the Lane--Emden equation in the complex plane. We thus find polytropic spherical shells providing hosting orbits to planets. On the basis of this model, we develop a numerical method which has three versions. In its three-dimensional version, the method is effective for systems with substantial uncertainties in the observed host star radius, and in the orbit of a particular planet (compared to the uncertainties in the orbits of the other planets); the method uses as fixed entry values the observed orbits of the remaining planets. In its two-dimensional version, the method is effective for systems with substantial uncertainty in the host star radius; in this case, the method uses as fixed entry values the observed orbits of the planets. The one-dimensional version was previously developed and applied to several systems; in this version, the observed values of the host star radius and of the planetary orbits are taken as fixed entry values. Our method can compute optimum values for the polytropic index of the global polytropic model which simulates the exoplanetary system, for the orbits of the planets, and (excluding the one-dimensional version) for the host star radius.   
\end{abstract}

\section{Introduction}
\label{intro}
Planetary orbits in the solar system and in exoplanetary systems have been studied by several authors within the framework of classical mechanics (see e.g. \citep{CK07}-\citep{PPL08}). Alternatively, other investigators use the frameworks of scale relativity, relativity theory regarding the finite propagation speed of gravitational interaction, and quantum mechanics (see e.g. \citep{HSG98}-\citep{OMC04}). 

We develop here a numerical method based on the equations of hydrostatic equilibrium of classical mechanics. These equations lead to the Lane--Emden differential equation, solved in the complex plane by the the so-called ``complex plane strategy'' (see e.g.~\citep{GKAR14}, Section~3), which is effective for numerically studying several astrophysical problems (see e.g. \citep{G88},~\citep{GV12}). The solution is the complex Lane--Emden function. According to the so-called ``global polytropic model'' (for preliminary concepts regarding this model, see \citep{GVD14}, Sections~2-3), polytropic spherical shells defined by succesive roots of  the real part  of the Lane--Emden function are appropriate places for accomodating planets.

Our method has three versions. The ``one-dimensional version'', developed earlier, has been applied to several exoplanetary systems (\citep{G14}-\citep{Ger17}). This version uses fixed entry values for the host star radius and for the planetary orbits. Entry values for the polytropic index of the global polytropic model which simulates the system are taken from a properly defined interval of values. This version can compute optimum values for the polytropic index and for the planetary orbits. 

The ``two-dimensional version'' is developed and used here for the first time. It takes fixed entry values for the planetary orbits. Entry values for the polytropic index of the global polytropic model and for the host star radius are taken from two properly defined intervals of values. The method can compute optimum values for the polytropic index, for the radius of the host star, and for the planetary orbits. 

The ``three-dimensional version'' is also developed and used here for the first time. It works with fixed entry values for the orbits of the planets, except for a particular planet with substantial uncertainty in its orbit in comparison with the uncertainties in the orbits of the other planets. Entry values for the polytropic index of the global polytropic model, for the host star radius, and for the orbit of the particular planet are taken from three properly defined intervals of values. The method can compute optimum values for the polytropic index, for the radius of the host star, and for the planetary orbits.

\section{The Method}
\label{3DM}
For convenience, we will use hereafter the definitions and symbols adopted in \cite{GVD14}.

The real part $\bar{\theta}(\xi)$ of the complex function $\theta(\xi)$ has a first root at $\xi_1 = \bar{\xi_1} + i \, \breve{\xi_0}$, a second root at $\xi_2 = \bar{\xi_2} + i \, \breve{\xi_0}$ with $\bar{\xi_2} > \bar{\xi_1}$, a third root at $\xi_3 = \bar{\xi_3} + i \, \breve{\xi_0}$ with $\bar{\xi_3} > \bar{\xi_2}$, etc. The polytropic sphere of polytropic index $n$ and radius $\bar{\xi}_1$ is the central component of a resultant polytropic configuration with further components the polytropic spherical shells $S_2$, $S_3$, \dots, defined by the pairs of radii $(\bar{\xi}_1, \, \bar{\xi}_2)$, $(\bar{\xi}_2, \, \bar{\xi}_3)$, \dots, respectively.
Each polytropic shell can be considered as an ideal hosting place for a planet. The most appropriate orbit radius $\bar{\Xi}_j \in [\bar{\xi}_{j-1},\,\bar{\xi}_j]$ is that at which $|\bar{\theta}|$ takes its maximum value inside $S_j$, 
\begin{equation}
\mathrm{max}|\bar{\theta}[S_j]| = |\bar{\theta}(\bar{\Xi}_j + i \, \breve{\xi}_0)|.
\label{maxth} 
\end{equation}
There are two further proper orbits with radii $\bar{\Xi}_\mathrm{Lj}$ and $\bar{\Xi}_\mathrm{Rj}$, such that 
\begin{equation}
\bar{\Xi}_\mathrm{Lj} < \bar{\Xi}_j < \bar{\Xi}_\mathrm{Rj}, 
\end{equation}
at which $|\bar{\theta}|$ becomes equal to its average value inside $S_j$,
\begin{equation}
\mathrm{avg}|\bar{\theta}[S_j]| = |\bar{\theta}(\bar{\Xi}_\mathrm{Lj} + i \, \breve{\xi}_0)|
                                = |\bar{\theta}(\bar{\Xi}_\mathrm{Rj} + i \, \breve{\xi}_0)|.
\label{avgth}   
\end{equation}
Accordingly, up to three planets can be hosted in $S_j$ on orbits with radii  $\bar{\Xi}_\mathrm{Lj}$, $\bar{\Xi}_\mathrm{j}$, and $\bar{\Xi}_\mathrm{Rj}$.

Our method can be applied to a system with $N_\mathrm{P}$ planets,  
\begin{equation}
\{P_m\} = \{P_m, \, m=1,\dots,N_\mathrm{P}\},
\label{system}
\end{equation}
and with corresponding observed distances from the host star  
\begin{equation}
\{A_m\} = \{A_m, \, m=1,\dots,N_\mathrm{P}\}, \ \ \ \mathrm{such~that} \ \ 
         A_{1} < A_{2} < \dots < A_{{N_\mathrm{P}}}.
\label{distances}
\end{equation}

The method is based on an algorithm which takes action over a three-dimensional parametric space 
\begin{equation}
\frak{S} = (\alpha_{\mathrm{[p]}i}, R_j, n_k), 
\end{equation}
where $\alpha_{\mathrm{[p]}i}$ are entry values for the orbit radius of a particular planet p with substantially large uncertainty in its observed value in comparison with the uncertaities in the orbit radii of the other planets; $R_j$ are entry values for the host star radius $R$; and $n_k$ are entry values for the polytropic index $n$ of the global polytropic model which simulates the system. 

It is expected that appropriate values of the polytropic index $n$ for modeling the planetary systems under consideration are about $n \sim 3$ (see e.g. \citep{Hor04}, Section~6.1 and references therein; see also \citep{GVD14}, Section~3 and references therein). In this study, entry values for $n$ are provided by an array
\begin{equation} 
\{n_k\} = \{n_k,k=1,\dots,N_\mathrm{n}+1\}
\label{nkarray}
\end{equation}
with elements  
\begin{equation}
n_k = 2.400 + 0.001 \times (k-1), \qquad k = 1,2,\dots,N_\mathrm{n}+1,
\label{Nn-now}
\end{equation} 
and 
\begin{equation}
N_\mathrm{n}=900.
\label{N1000}
\end{equation}

The 901 complex ``initial value problems'' (IVP, IVPs) counted in Eq.~(\ref{Nn-now}) are solved by the Fortran package \texttt{dcrkf54.f95} \citep{GV12} which is a Runge--Kutta--Fehlberg code of fourth and fifth order modified for solving complex IVPs established on ordinary differential equations of various complex functions in one complex variable, along contours prescribed as continuous chains of straight-line segments; details on the usage of \texttt{dcrkf54.f95} are also given in \citep{GVD14} (Section~4). Integrations proceed along the contour  
\begin{equation}
\frak{C} = \{\xi_0 = (10^{-4},\,10^{-4}) \rightarrow 
                               \xi_\mathrm{end} = (10^5,\,10^{-4})
                               \}.
\label{CJSUN}
\end{equation}
This contour belongs to the special form~(8) of \citep{GVD14}; various contours and their characteristics are defined in \cite{GV12} (Section~5).

Since physical interest focuses on real parts of complex orbit radii, we will hereafter quote only such values and, for simplicity, we will drop overbars denoting real parts.

For each $n_k$, the algorithm computes the array $\{\xi_l,l=1,\dots,N_\mathrm{r}\}_{n_k}$, where the integer $N_\mathrm{r}$ is chosen adequately large. Thus, next to the first root
\begin{equation}
\{(\xi_1)\}_{n_k}=\{(\xi_1)_k,k=1,\dots,N_\mathrm{n}\}, 
\label{firstroot}
\end{equation}
there are computed $N_\mathrm{r}-1$ roots
with respective $N_\mathrm{r}-1$ hosting shells, and 
\begin{equation}
N_\mathrm{H}=3 \, (N_\mathrm{r}-1)
\label{NH} 
\end{equation}
hosting orbits; namely,
\begin{equation}
\left\{\Psi_l,l=1,\dots,N_\mathrm{H}\right\}_{n_k} =
\{\Xi_\mathrm{L2},\Xi_\mathrm{2},\Xi_\mathrm{R2},\dots,
  \Xi_\mathrm{LN_\mathrm{H}},\Xi_\mathrm{N_\mathrm{H}},\Xi_\mathrm{RN_\mathrm{H}}\}_{n_k}.
\label{hostingorbits-1d}
\end{equation}
Accordingly, computations over all entry values $n_k$ give the two-dimensional array 
\begin{equation} 
\left\{\Psi_{kl}\right\} = \left\{\Psi_{kl}, \, k=1,\dots,N_\mathrm{n}+1, \, l=1,\dots,N_\mathrm{H}\right\}.
\label{hostingorbits-2d}
\end{equation}

Entry values for the host star radius are taken from the array 
\begin{equation}
\{R_j\} = \{R_j,j=1,\dots,N_\mathrm{R}+1\}.
\label{radii}
\end{equation}
In particular, if the quoted host radius (observed or estimated by an appropriate model) is $R_\mathrm{q}$  with uncertainty $\pm (R_\mathrm{q})_\mathrm{u}$, then the array elements $R_j$ are given by
\begin{equation}
R_j = R_\mathrm{q}-(R_\mathrm{q})_\mathrm{u} + (j-1) \ \frac{2 \, (R_\mathrm{q})_\mathrm{u}}{N_\mathrm{R}}, \qquad
j=1,\dots,N_\mathrm{R}+1.
\label{hostradii}
\end{equation}    
 
Entry values for the orbit radius of a particular planet p with substantially large uncertainty in its orbit radius, compared to those of the other planets, are provided by the array
\begin{equation}
\{\alpha_{\mathrm{[p]}i}\} = \{\alpha_{\mathrm{[p]}i},i=1,\dots,N_\alpha+1\}. 
\label{alphas}
\end{equation}
In detail, if the quoted observed distance of the planet p is $A_\mathrm{[p]q}$ with  uncertainty $\pm (A_\mathrm{[p]q})_\mathrm{u}$, then the array elements $\alpha_{\mathrm{[p]}i}$ are given by 
\begin{equation}
\alpha_{\mathrm{[p]}i} = A_\mathrm{[p]q}-(A_\mathrm{[p]q})_\mathrm{u} +  
(i-1) \ \frac{2 \, (A_\mathrm{[p]q})_\mathrm{u}}{N_\alpha}, \\\
i=1,\dots,N_\alpha+1.
\label{appi}
\end{equation} 

For the model with  current indices $i$, $j$, $k$, the method assigns to the planet $P_m$ the orbit radius $\alpha_m$ for which the ratio $\alpha_{ml} = \Psi_{kl}/(\xi_1)_k$ has the minimum absolute percent difference with respect to the ratio $A_m/R_j$ among all the indices $l$.
Next, the method computes the sum of the minimum absolute percent differences of the assigned orbits over all the planets of the system. 

Among the $(N_\alpha+1) \times (N_\mathrm{R}+1) \times (N_\mathrm{n}+1)$ resolved  models, ``optimum model'', with indices  $I$, $J$, $K$, is the model for which: (a) the sum of the minimum absolute percent differences becomes minimum among all  the models (denoted by $\Delta_\mathrm{opt}$), and (b) the indices $I$ and $J$ are not coinciding with their starting values, $i=1$ and/or $j=1$, or with their terminating values,  $i=(N_\alpha+1)$ and/or $j=(N_\mathrm{R}+1)$. 
The meaning of the condition (b) is that the observed uncertainties involved in the definitions of the arrays~(\ref{hostradii}) and~(\ref{appi}) of the entry values should indeed bound the host star radius and the orbit radius of the particular planet p. Failure in obeying this condition points out that the observed uncertainties have been probably underestimated. In such a case, the array intervals defined by the relations~(\ref{hostradii}) and~(\ref{appi}) are properly extended, either to the left or to the right, and the computations are repeated for these new intervals. 

The polytropic index $n_K$ is the ``optimum polytropic index'' for the global polytropic model simulating the system; the host star radius $R_J$ is the ``optimum host star radius'' predicted by the method; and the orbit radius $\alpha_{\mathrm{[p]}I}$ is the ``optimum orbit radius'' for the particular planet p predicted by the method.

\section{The Fortran Code}
\label{fortran}
Our method is implemented by a Fortran code consisting of two packages. The first package treats a particular system up to Eq.~(\ref{hostingorbits-2d}) by solving the IVPs involved in the problem. Basic constituent of the first package is the Fortran code \texttt{dcrkf54.f95} \citep{GV12}. The use of this code has been adequately described in previous investigations (\citep{GV12}, \citep{GVD14}). In fact, the first package performs the bookkeeping of the numerical results computed by \texttt{dcrkf54.f95}. The second package controls all necessary iterations  (DO~Loops) over the entry values related to Eqs.~(\ref{nkarray})-(\ref{Nn-now}), (\ref{radii})-(\ref{hostradii}), and (\ref{alphas})-(\ref{appi}).

\subsection{Macro-description of the second package}
\label{2pack}
To proceed with a macro-description of the second package by using Fortran conventions, we need to assign Fortran names to the following involved variables: \\    
\\    
\texttt{N\_\,P}=$N_\mathrm{P}$ (Eq.~(\ref{system})), \\
\texttt{A\_\,P(1:N\_\,P})=$\{A_m,m=1,\dots,N_\mathrm{P}\}$ (Eq.~(\ref{distances})), \\
\texttt{N\_\,n}=$N_\mathrm{n}+1$ (Eq. (\ref{nkarray})), \\
\texttt{PLI(1:N\_\,n})=$\{n_k,k=1,\dots,N_\mathrm{n}+1\}$ (Eq.~(\ref{nkarray})), \\
\texttt{x1(1:N\_\,n})=$\{(\xi_1)_k,k=1,\dots,N_\mathrm{n}+1\}$
(Eq.~(\ref{firstroot})), \\ 
\texttt{N\_\,H}=$N_\mathrm{H}$ (Eq.~(\ref{NH})), \\
\texttt{PSI(1:N\_\,n,1:N\_\,H})=$\{\Psi_{kl},k=1,\dots,N_\mathrm{n}+1,\,
l=1,\dots,N_\mathrm{H}\}$ (Eq.~(\ref{hostingorbits-2d})), \\
\texttt{N\_\,R}=$N_\mathrm{R}+1$ (Eq.~(\ref{radii})), \\
\texttt{R(1:N\_\,R})=$\{R_j,j=1,\dots,N_\mathrm{R}+1\}$ (Eq.~(\ref{radii})), \\
\texttt{Position\_\,of\_\,App}=$N_\mathrm{[p]}$ (Eqs.~(\ref{alphas})-(\ref{appi})), \\
\texttt{N\_\,a}=$N_\alpha+1$ (Eq.~(\ref{alphas})), \\ \texttt{app(1:N\_\,a})=$\{\alpha_{\mathrm{[p]}i},i=1,\dots,N_\alpha+1\}$  (Eq.~(\ref{alphas})). \\
\texttt{SUM\_\,min\_opt}=$\Delta_\mathrm{opt}$ (Section~\ref{3DM}) \\
\\
Then a macro-description of the second package has as follows:     
\begin{footnotesize}
\begin{verbatim}
! Start of Loop AppOrbitRadii
      AppOrbitRadii: DO I=1,N_a
! Start of Loop HostStarRadii 
          HostStarRadii: DO J=1,N_R
! Start of Loop PolytropicIndices
              PolytropicIndices: DO K=1,N_n
! Start of Loop HostingOrbits 
                  HostingOrbits: DO L=1,N_H
                      alpha(L)=PSI(K,L)/x1(K)
                  END DO HostingOrbits          ! End of Loop HostingOrbits                                                  
! Start of Loop HostedPlanets 
                  HostedPlanets: DO M=1,N_P
                      IF (M==Position_of_App) THEN
                          A_P(M)=app(I)
                      END IF
                      AUX_A_P(1:N_H)=A_P(M)/R(J)
                      difmin_P(M)=MINVAL(ABS(AUX_A_P-alpha))
                      MINPOS_M   =MINLOC(ABS(AUX_A_P-alpha))
                      orbits_P(M)=MINPOS_M
                  END DO HostedPlanets          ! End of Loop HostedPlanets                                                  
                  SUM_min_n(K)     =SUM(difmin_P(1:N_P))
                  orbits_n(K,1:N_P)=orbits_P(1:N_P)
              END DO PolytropicIndices          ! End of Loop PolytropicIndices                                                
              SUM_min_R(J)         =MINVAL(SUM_min_n(1:N_n))
              MINPOS_J             =MINLOC(SUM_min_n(1:N_n))
              PLI_opt_R(J)         =PLI(MINPOS_J) 
              ORBITS_opt_R(J,1:N_P)=orbits_n(MINPOS_J,1:N_P)
          END DO HostStarRadii                  ! End of Loop HostStarRadii                                                  
          SUM_min_a(I)         =MINVAL(SUM_min_R(1:N_R))
          MINPOS_I             =MINLOC(SUM_min_R(1:N_R))
          R_opt_a(I)           =R(MINPOS_I) 
          PLI_opt_a(I)         =PLI_opt_R(MINPOS_I)
          ORBITS_opt_a(I,1:N_P)=ORBITS_opt_R(MINPOS_I,1:N_P) 
      END DO AppOrbitRadii                      ! End of Loop AppOrbitRadii                                                
! Final Session: Overall Estimates
      SUM_min_opt      =MINVAL(SUM_min_a(1:N_a))
      MINPOS           =MINLOC(SUM_min_a(1:N_a)) 
      app_opt          =app(MINPOS)
      RADIUS_opt       =R_opt_a(MINPOS)
      POLIND_opt       =PLI_opt_a(MINPOS) 
      ORBITS_opt(1:N_P)=ORBITS_opt_a(MINPOS,1:N_P)
\end{verbatim}
\end{footnotesize}

\subsection{Remarks on the DO loops and the final session}
\label{doloops}
Loop \texttt{HostingOrbits} 

For the current K-th polytropic index \texttt{PLI(K)}, the array \texttt{alpha(1:N\_\,H)} is set equal to the array subobject \texttt{PSI(K,1:N\_\,H)/x1(K)} having elements the orbit radii measured with unit the first root \texttt{x1(K)}. \\
\\
Loop \texttt{HostedPlanets} 

The elements of the auxiliary array \texttt{AUX\_\,A\_\,P(1:N\_\,H)} are set equal to the observed distance \texttt{A\_\,P(M)} of the current M-th planet measured with unit the current J-th host's radius \texttt{R(J)}. If \texttt{M=Position\_\,of\_\,App}, then \texttt{A\_\,P(M)} is set equal to the current I-th entry \texttt{app(I)}. \texttt{MINPOS\_\,M} is the position in the array  \texttt{alpha(1:N\_\,H)} occupied by the orbit radius having the minimum absolute percent diference \texttt{difmin\_\,P(M)} relative to the distance \texttt{A\_\,P(M)} of the current M-th planet, and the polytropic orbit  \texttt{orbits\_\,P(M)} hosting this planet is set equal to \texttt{MINPOS\_\,M}. \\
\\
Loop \texttt{PolytropicIndices}

For the current K-th polytropic index \texttt{PLI(K)}, the element \texttt{SUM\_\,min\_\,n(K)} is set equal to the sum of the elements of the array \texttt{difmin\_\,P(1:N\_\,P)}. Next, the array subobject \texttt{orbits\_\,n(K,1:N\_\,P)} is set equal to the array \texttt{orbits\_\,P(1:N\_\,P)}; so, the rank-2 array \texttt{orbits\_\,n(1:N\_\,n,1:N\_\,P)} is the extension of the rank-1 array \texttt{orbits\_\,P(1:N\_\,P)} over the dimension \texttt{1:N\_\,n}. \\
\\
Loop \texttt{HostStarRadii} 

For the current J-th host's radius \texttt{R(J)}, the element \texttt{SUM\_\,min\_\,R(J)} is set equal to the minimum value of the array \texttt{SUM\_\,min\_\,n(1:N\_\,n)}, occupying the position \texttt{MINPOS\_\,J}. The element  \texttt{PLI\_\,opt\_\,R(J)} having the minimum sum \texttt{SUM\_\,min\_ \,R(J)} is set equal to the element \texttt{PLI(MINPOS\_\,J)} of the array \texttt{PLI(1:N\_\,n)}. The array subobject \texttt{ORBITS\_\,opt\_\,R(J,1:N\_\,P)} is set equal to the array subobject \texttt{orbits\_\,n(MINPOS\_\,J,1:N\_\,P)}; so, the rank-2 array  \texttt{ORBITS\_\,opt\_\,R(1:N\_\,R,1:N\_\,P)} is the extension of  \texttt{ORBITS\_\,opt\_\,R(J,1:N\_\,P)} over the dimension \texttt{1:N\_\,R}. \\
\\
Loop \texttt{AppOrbitRadii}

For the current I-th entry \texttt{app(I)}, the element \texttt{SUM\_\,min\_\,a(I)} is set equal to the minimum value of the array \texttt{SUM\_\,min\_\,R(1:N\_\,R)}, occupying the position \texttt{MINPOS\_\,I}. The element  \texttt{R\_\,opt\_\,a(I)} having the minimum sum \texttt{SUM\_\,min\_\,a(I)} is set equal to the element \texttt{R(MINPOS\_\,I)} of the array \texttt{R(1:N\_\,R)}. The element  \texttt{PLI\_\,opt\_\,a(I)} having the minimum sum \texttt{SUM\_\,min\_\,a(I)} is set equal to the element \texttt{PLI\_\,opt\_R(MINPOS\_\,I)} of the array \texttt{PLI\_\,opt\_\,R(1:N\_\,R)}. The array subobject \texttt{ORBITS\_\,opt\_\,a(I,1:N\_\,P)} is set equal to the array subobject \texttt{ORBITS\_\,opt\_\,R( MINPOS\_\,I, 1:N\_\,P)}; so, the rank-2 array  \texttt{ORBITS\_\,opt\_\,a(1:N\_\,a,1:N\_\,P)} is the extension of \texttt{ORBITS\_\,opt\_\,a(I,1:N\_\,P)} over the dimension \texttt{1:N\_\,a}. \\
\\
Final Session: Overall Estimates

The ``optimum minimum sum'', \texttt{SUM\_\,min\_opt}, of the absolute percent differences of the computed planetary orbit radii relative to their observed values  is the minimum value of the array \texttt{SUM\_\,min\_\,a(1:N\_\,a)}, occupying the position \texttt{MINPOS}. The ``optimum orbit radius'', \texttt{app\_\,opt}, of a particular planet p with a substantially high uncertainty (compared to the uncertainties in the distances of the other planets)  is the element \texttt{app(MINPOS)} of the array \texttt{app(1:N\_\,a)}. The ``optimum host star radius'', \texttt{RADIUS\_\,opt}, is the element \texttt{R\_\,opt\_\,a(MINPOS)} of the array \texttt{R\_\,opt\_\,a(1:N\_\,a)}. The ``optimum polytropic index'' \texttt{POLIND\_\,opt} is the element \texttt{PLI\_\,opt\_\,a(MINPOS)} of the array \texttt{PLI\_\,opt\_\,a(1:N\_\,a)}. The ``optimum planetary orbits'', \texttt{ORBITS\_\,opt(1:N\_\,P)}, are the respective elements of the array \texttt{ORBITS\_\,opt\_\,a(MINPOS,1:N\_\,P)}. \\

\section{The Two- and One-Dimensional Versions}
\label{2DM}
There are numerous exoplanetary systems listed in NExA, as well as in other exoplanet archives, for which the uncertainties in the orbits of their planets are comparable. In addition, for several exoplanetary systems appearing in the archives, observational data regarding uncertaities in the planetary orbits are missing. In both cases, we cannot distinguish a planet with a substantially larger uncertainty in its orbit. Hence, the two-dimensional version of our method is the effective one for such systems.  
It is easy to implement this version, instead of its three-dimensional counterpart,  by simply setting $N_\alpha=0$ in the relation~(\ref{appi}) and $\alpha_{\mathrm{[p]}1} = A_1$ for the unique entry value.

In this paper, the one-dimensional version of our method is not implemented. Typically, this version is appropriate for studying systems with small or missing observational uncertainties in the radii of the host stars and, as said above, with comparable or  missing uncertainties in the planetary orbits. Implementation of this version can be achieved by additionally setting $N_\mathrm{R}=0$ in the relation~(\ref{hostradii}) and $R_1=R_\mathrm{q}$ for the unique entry value. Numerical results of several exoplanetary systems computed by the one-dimensional method are given in \citep{G14}-\citep{Ger17}; some of these results are discussed below.

In terms of the quantities involved in the Fortran code (Section~\ref{doloops}), use of the two-dimensional version is achieved by setting \texttt{N\_\,a=1} and \texttt{app(1)=A\_\,P(1)}. To apply the one-dimensional version, we additionally set \texttt{N\_\,R=1} and \texttt{R(1)=R\_q}.

\section{Numerical Results}
\label{results}
We select the exoplanetary systems Kepler-11, Kepler-90, Kepler-215, HD~10180, HD~34445, and TRAPPIST-1 as paradigms for applying our method. Relevant observational data are included in the ``NASA Exoplanet Archive'' (https://exo-planetarchive.ipac.caltech.edu/ --- hereafter abbreviated ``NExA'') unless explicitly stated otherwise.
The systems TRAPPIST-1 and Kepler-90 are the only ones in NExA with number of planets $N_\mathrm{P} \geq 7$. Next, among the six systems listed in NExA with $N_\mathrm{P}=6$, we have selected three of them: Kepler-11, HD~10180, and HD~34445. On the other hand, the system Kepler-215 has four planets, the uncertainties in their orbits are missing from NExA, and the uncertainty in the host star radius is substantially large when compared to the uncertainties of other cases. The numerical results for the selected paradigms reveal some interesting aspects of the method, verifying in turn its flexibility and reliabity (to be discussed below). 

The symbols involved in Tables~\ref{k11}-\ref{trappist1} have the following meaning: $n_\mathrm{opt}$ is the optimum polytropic index for the global polytropic model which simulates the exoplanetary system. $\xi_{1\mathrm{opt}}$ is the optimum radius of the host star given in classical polytropic units, in which the length unit is equal to the polytropic parameter $\alpha$ (\citep{GVD14}, Eq.~(3b)). $R_\mathrm{opt}$ is the optimum radius of the host star expressed in solar radii ($R_\odot$). $R_\mathrm{q}$ is the quoted radius of the host star, either observed or computed by a model, and $(R_\mathrm{q})_\mathrm{u}$ is its uncertainty, both given in solar radii.
Shell radii and orbit radii are given in astronomical units (AU). For successive shells $S_j$ and $S_{j+1}$, inner radius of $S_{j+1}$ is the outer radius of $S_j$. Percent differences $\%D_j$ in the computed orbit radii $\alpha_j$ are given with respect to the corresponding distances $A_j$, $\%D_j = 100 \times |(A_j - \alpha_j)| / A_j$. Parenthesized signed integers denote powers of 10. 
 
Hereafter, radii of the host stars and their uncertainties will be expressed in solar radii ($R_\odot$); planetary orbit radii and their uncertainties will be expressed in astronomical units (AU). Furthermore, the meaning assigned to the term ``difference'' will be that of ``absolute percent difference''.

\subsection{The System Kepler-11}
\label{sk11}
Regarding the 6-planet system Kepler-11 (see e.g.~\citep{L13}-\citep{BBM17}), 
the computed optimum minimum sum $\Delta_\mathrm{opt}$ is found to be  
\begin{equation}
\Delta_\mathrm{opt} \simeq 6.7\%.
\end{equation}
The average difference in the computed distances of the six planets is $\simeq 1.1\%$.
Smaller difference is that for g's distance, $\simeq 0.009\%$. Larger difference is that for d's distance, $\simeq 4.1\%$. 

The shell No~5 is occupied by two planets: b and c. The former is resident of the ``maximum-density orbit'' (Eq.~(\ref{maxth})) with radius $\alpha_\mathrm{b} = \alpha_5$; and the latter is hosted on the ``average-density orbit'' (Eq.~(\ref{avgth})) with radius $\alpha_\mathrm{c} = \alpha_\mathrm{R5}$. Likewise, the shell No~6 is occupied by the planets d and e. The former is resident of the average-density orbit with radius $\alpha_\mathrm{d} = \alpha_\mathrm{L6}$; the latter is resident of the average-density orbit with radius $\alpha_\mathrm{e} = \alpha_\mathrm{L6}$.

The computed optimum radius $R_\mathrm{opt}$ for the star Kepler-11 lies to the left of the interval $[1.043,1.082]$ determined by the uncertainty $(R_\mathrm{q})_u = _{-0.022}^{+0.017}$ in the quoted radius $R_\mathrm{q} = 1.065$, 
\begin{equation}
R_\mathrm{opt} = 1.0115 < 1.043,
\end{equation}  
and its absolute percent difference relative to $R_\mathrm{q}$ is 
\begin{equation}
\%D(R_\mathrm{opt}) \simeq 5.1\%.
\end{equation}
On the other hand, however, in \citep{GFG17} (Sect.~3) the revised value $1.021$ is assigned to the radius of the star Kepler-11, with an uncertainty $\pm 0.025$. Therefore, the revised interval becomes $[0.996,1.046]$ and thus the computed optimum radius lies in this interval,
\begin{equation}
R_\mathrm{opt} = 1.0115 \in [1.021 \pm 0.025],
\end{equation}  
with a difference $\simeq 0.93\%$ relative to the revised radius.

It is worth mentioning here that in \citep{G14} (Eq.~(8) and Table~5) we applied the one-dimensional method to the system Kepler-11, with fixed radius $1.065$ for the host star, and we found $n_\mathrm{opt}=2.779$ and $\Delta_\mathrm{opt} \simeq 32.7\%$.

\subsection{The System Kepler-90 (KOI-351)}
\label{sk90}
For the 8-planet system Kepler-90 (see e.g. \citep{CCL13}-\citep{SV17}),
there is no information in NExA for the orbit radius $A_\mathrm{i}$ of the planet i; concerning this distance, we adopt from \citep{SV17} (Table\,5) the value $A_\mathrm{i} = 0.20277$.  

For the optimum model we find  
\begin{equation}
\Delta_\mathrm{opt} \simeq 10.7\%.
\end{equation}
The average difference in the computed distances of the eight planets is $\simeq 1.3\%$. Smaller difference is that for d's distance, $\simeq 0.003\%$. Larger difference is that for i's distance, $\simeq 4.9\%$. 

Table \ref{k90} shows that each of the eight shells 4-11 is occupied by only one planet. The planets i and d occupy maximum-density orbits within their hosting shells No~6 and No~7, respectively. The other planets of the system occupy either left or right average-density orbits within their shells. 

The optimum radius $R_\mathrm{opt}$ for the star Kepler-90 lies in the interval determined by the uncertainty $(R_\mathrm{q})_u$ of the quoted radius $R_\mathrm{q}$,
\begin{equation}
R_\mathrm{opt} = 1.2073 \in [1.2 \pm 0.1],
\end{equation}  
and its difference relative to $R_\mathrm{q}$ is equal to
\begin{equation}
\%D(R_\mathrm{opt}) \simeq 0.61\%.
\end{equation}

Note that in \citep{G14} (Eq.~(10) and Table~9) we studied the system Kepler-90 with the one-dimensional method by taking fixed radius $1.2$ for the host star, and we computed the values $n_\mathrm{opt}=2.819$ and $\Delta_\mathrm{opt} \simeq 13.4\%$.

\subsection{The System Kepler-215}
\label{sk215}
For the 4-planet system Kepler-215 (see e.g.~\citep{RBM14}), the computed optimum model gives
\begin{equation}
\Delta_\mathrm{opt} \simeq 1.6\%.
\end{equation}
The average difference in the computed distances of the four planets is $\simeq 0.39\%$. Smaller difference is that for b's distance, $\simeq 0.008\%$. Larger difference is that for e's distance, $\simeq 0.96\%$. Table\,\ref{k215} shows that each of the four shells 4-7 is hosting only one planet. The planets occupy the corresponding maximum-density orbits within their shells, with the exception of the planet b which occupies the left average-density orbit in the shell No~4. 

The computed optimum radius $R_\mathrm{opt}$ for the star Kepler-215 lies in the interval determined by the uncertainty $(R_\mathrm{q})_u$ of the quoted radius $R_\mathrm{q}$,
\begin{equation}
R_\mathrm{opt} = 1.0253 \in [1.027 \pm 0.236],
\end{equation}  
and its difference relative to $R_\mathrm{q}$ is equal to
\begin{equation}
\%D(R_\mathrm{opt}) \simeq 0.17\%.
\end{equation}

\subsection{The System HD~10180}
\label{shd10180}
For the 6-planet system HD~10180 (see e.g.~\citep{LSM10}-\citep{KG14}), the computed optimum model gives
\begin{equation}
\Delta_\mathrm{opt} = 14.2\%.
\end{equation}
The average difference in the computed distances of the six planets is $\simeq 2.4\%$.
Smaller difference is that for d's distance, $\simeq 0.003\%$. Larger difference is that for c's distance, $\simeq 7.3\%$. 
 
Table \ref{hd10180} reveals that each of the six polytropic shells 3-6, 8, and 11 is hosting only one planet. The planets occupy the maximum-density orbits within their shells, except for the planets c and g which occupy the left average-density orbits within their shells No~3 and No~8, respectively. 

The optimum radius $R_\mathrm{opt}$ for the star HD~10180 lies in the interval determined by the uncertainty $(R_\mathrm{q})_u$ of the quoted radius $R_\mathrm{q}$,
\begin{equation}
R_\mathrm{opt} = 1.0898 \in [1.109 \pm 0.036],
\end{equation}  
and its difference relative to $R_\mathrm{q}$ is equal to
\begin{equation}
\%D(R_\mathrm{opt}) \simeq 1.7\%.
\end{equation} 

It is interesting to mention here that in \citep{G15b} (Eq.~(1) and remarks following this equation) we studied the system HD~10180 by applying the one-dimensional method, taking fixed radius $1.2$ for the host star. Our computations resulted in the values $n_\mathrm{opt}=3.060$ and $\Delta_\mathrm{opt} \simeq 45.2\%$.

\subsection{The System HD~34445}
\label{shd34445}
Concerning the 6-planet system HD~34445 (see e.g. \citep{HJM10}-\citep{VBB17}), it is apparent from the available data that the planet g has a substantially larger uncertainty, $(A_\mathrm{[g]q})_\mathrm{u} = \pm 1.02$, in its observed distance, $A_\mathrm{[g]q} = 6.36$, in comparison with the uncertainties in the distances of the other five planets of the system. Thus, the three-dimensional version of our method is the effective one for this system. 

For the optimum model, we find 
\begin{equation}
\Delta_\mathrm{opt} \simeq 0.96\%.
\end{equation}
So, the average difference in the computed distances of the six planets is $\simeq 0.16\%$. Smaller difference is that for e's distance, $\simeq 0.003\%$ (the zero difference for g's distance is excluded from the comparison, since this distance plays a parametric role in our method). Larger difference is that for c's distance, $\simeq 0.35\%$.  

Table \ref{hd34445} shows that each of the six shells 9, 12, 14, 20, 24, 41, is hosting only one planet. The planets occupy left or right average-density orbits within their shells, with the exception of the planet e which occupies the maximum-density orbit of the shell No~9.

The computed optimum radius $R_\mathrm{opt}$ for the star HD 34445 lies to the left of the interval $[1.36,1.40]$ determined by the uncertainty $(R_\mathrm{q})_u = \pm 0.02$ in the quoted radius $R_\mathrm{q} = 1.38$, that is
\begin{equation}
R_\mathrm{opt} = 1.3400 < 1.36,
\end{equation}  
and its difference relative to $R_\mathrm{q}$ is equal to
\begin{equation}
\%D(R_\mathrm{opt}) \simeq 2.9\%.
\end{equation}
However, according to the data given in \citep{HJM10} (Table~1; see also \citep{VBB17}, Table~1), the uncertainty in the radius of HD~34445 is assigned the value $\pm 0.08$; so, the interval of values becomes $[1.30,1.46]$ and the computed optimum radius lies in this interval,
\begin{equation}
R_\mathrm{opt} = 1.3400 \in [1.38 \pm 0.08].
\end{equation}  

Regarding the optimum distance $a_\mathrm{[g]opt}$ of the planet g, we find that it lies in the interval defined by the uncertainty $(A_\mathrm{[g]q})_\mathrm{u}$ of the quoted distance $A_\mathrm{[g]q}$,
\begin{equation} 
a_\mathrm{[g]opt} = 6.5002 \in [6.36 \pm 1.02]
\end{equation} 
with a difference relative to $A_\mathrm{[g]q}$ equal to 
\begin{equation}
\%D(a_\mathrm{[g]opt}) \simeq 2.2\%.
\end{equation}

\subsection{The System TRAPPIST-1}
\label{stp1}
For the 7-planet system TRAPPIST-1 (see e.g.~\citep{CK19c}, \citep{L17}-\citep{D18}), there is no information in NExA  for the orbit radius $A_\mathrm{h}$ of the planet h; for this orbit radius, we adopt from \citep{G17} (Table~1) the  value $A_\mathrm{h} = 0.063$. 

From the available data (\citep{G17}, Table~1), we verify that the planet h has a  substantially larger uncertainty, $(A_\mathrm{[h]q})_\mathrm{u} = _{-0.013}^{+0.027}$, in its observed distance, $A_\mathrm{[h]q} = 0.063$, in comparison with the uncertainties in the distances of the other six planets of the system. Thus, the three-dimensional version of our method is the effective one for this system. 

For the computed optimum model, the optimum minimum sum $\Delta_\mathrm{opt}$ is   
\begin{equation}
\Delta_\mathrm{opt} \simeq 6.3\%,
\end{equation}
and the average difference in the computed distances of the seven planets is $\simeq 0.9\%$. Smaller difference is that for d's distance, $\simeq 0.06\%$ (the zero difference for h's distance is excluded from the comparison, since this distance plays a parametric role in our method). Larger difference is that for g's distance, $\simeq 1.9\%$.
  
Table \ref{trappist1} shows that each of the seven shells 6-11,~13 is hosting only one planet. The planets occupy left or right average-density orbits in their shells, with the exception of the planets c and h which occupy the maximum-density orbits within their shells No~7 and No~13, respectively.

The computed optimum radius $R_\mathrm{opt}$ for the star TRAPPIST-1 lies beyond the right bound $0.1206$ of the interval $[0.1134,0.1206]$ determined by the uncertainty $(R_\mathrm{q})_u = \pm 0.0036$ in the quoted radius $R_\mathrm{q} = 0.117$, that is
\begin{equation}
R_\mathrm{opt} = 0.1228 > 0.1206,
\end{equation}  
and its absolute percent difference relative to $R_\mathrm{q}$ is equal to
\begin{equation}
\%D(R_\mathrm{opt}) \simeq 4.9\%.
\end{equation}
It is worth remarking, however, that in Table~1 of \citep{GFG17} the updated value for the radius of the star TRAPPIST-1 is $0.121$ with an uncertainty $\pm 0.003$. Accordingly, the updated interval of values becomes $[0.118,0.124]$ and the computed optimum radius lies in this interval,
\begin{equation}
R_\mathrm{opt} = 0.1228 \in [0.121 \pm 0.003],
\end{equation}  
with a difference $\simeq 1.5\%$ relative to the updated radius.   

Regarding the optimum distance of the planet h, we find that it lies in the interval defined by the uncertainty $(A_\mathrm{[h]q})_\mathrm{u}$ of the quoted distance $A_\mathrm{[h]q}$,
\begin{equation} 
a_\mathrm{[h]opt} = 0.0635 \in [0.063 _{-0.013}^{+0.027}],
\end{equation} 
with a difference relative to $A_\mathrm{[h]q}$ equal to 
\begin{equation}
\%D(a_\mathrm{[h]opt}) \simeq 0.8\%.
\end{equation}

It would be useful to quote here a previous investigation (\citep{Ger17}; Eq.~(1) and Table~1), in which we treated numerically the system TRAPPIST-1 by applying the one-dimensional method, with fixed radius $0.117$ for the host star and fixed orbit radius $0.063$ for the planet h. Our computations resulted in the values $n_\mathrm{opt}=2.525$ and $\Delta_\mathrm{opt} \simeq 44.2\%$.

\section{Discussion}
\label{discussion}
The predictions given by our method can be eventually verified by future observations and/or new numerical models. An interesting case pointing to prediction(s) arises when the method  fails to satisfy the condition~(b)~(Section~\ref{3DM}) in its first run. As discussed in Section~\ref{3DM}, failure in fulfilling the condition~(b) shows that the corresponding observed uncertainties have been probably underestimated. If so, the method extends properly the intervals ~(\ref{hostradii}) and/or~(\ref{appi}) of entry values for the host star radius and/or for the orbit radius of a particular planet p with substantial uncertainty in its orbit, and then proceeds with a second run.  

Such a case has emerged during the study of the system Kepler-11. As said in Section~\ref{sk11}, the optimum host star radius given in Table~\ref{k11} lies to left of the interval determined by the observed uncertainties. This optimum value has been computed by a second run, with the interval of entry values properly extended to the left. By taking into account the revised value $1.021$ and its uncertainty $\pm 0.025$ quoted in \citep{BBM17} (Section~3), we verify that the computed optimum radius lies in the revised interval of values. 

Likewise, as said in Section~\ref{shd34445}, in the case of the system HD~34445 the optimum radius of the host star given in Table~\ref{hd34445} lies to the left of the interval determined by the observed uncertainties. This optimum radius has been computed by a second run of the method with the interval of entry values properly extended to the left. By taking into account the revised uncertainty $\pm 0.08$ in the radius of HD~34445 given in \citep{HJM10} (Table~1; see also \citep{VBB17}, Table~1), we deduce that the computed optimum radius lies in the revised interval of values.

A third similar case has emerged in the treatment of the system TRAPPIST-1. As said in Section~\ref{stp1}, the optimum host star radius given in Table~\ref{trappist1} lies to right of the interval determined by the observed uncertainties. This value has been computed by a second run, with interval of values properly extended to the right. In accordance with the updated value $0.121$ for the radius of the star TRAPPIST-1 and its uncertainty $\pm 0.003$ given in \citep{GFG17} (Table~1), the computed optimum radius lies in this updated interval of values.   

Furthermore, our method can show flexibility when studying systems with small or missing uncertainties in the radii of the host stars. In such systems, the method can `pretend' that there are certain appreciable uncertainties in the observed radii and, accordingly, resolve the systems by the two-dimensional version instead of its one-dimensional counterpart (which could typically used). A relevant case is that of the system Kepler-11. The quoted uncertainties in the radius of the host star are small (in fact, the smallest ones among the selected paradigms), $(R_\mathrm{q})_\mathrm{u}=_{-0.022}^{+0.017}$, i.e. $\sim (\pm 2\%)$, with respect to the quoted radius $R_\mathrm{q}=1.065$. As said above, the method has extended to the left the interval of entry values for the radius of the host star and, with a second run, has computed the optimum value $1.0115$ for the radius of the star Kepler-11. 

A general remark concerning previous numerical results derived by the one-dimensional method is that the minimum $\Delta_\mathrm{opt}$ is relatively large when compared to the corresponding values found by the two- and three-dimensional methods. For example, for the system Kepler-11 (Section~\ref{sk11}), this minimum is $\sim 5$ times greater than the value $\simeq 6.7\%$ estimated by the two-dimensional method; for the system HD~10180 (Section~\ref{shd10180}), this minimum is $\sim 3$ times greater than the value $\simeq 14.2\%$ estimated by the two-dimensional method; and for the system TRAPPIST-1 (Section~\ref{stp1}), the minimum is $\sim 7$ times greater than the value $\simeq 6.3\%$ estimated by the three-dimensional method. 

In addition, there is a relevant question arising here: Can we reduce `deliberately' the dimensions of our method in a particular application without decreasing the accuracy of the respective results? As discussed above, the astrophysical data available for a given exoplanetary system show themselves the need for applying to this system the three-dimensional method. Namely, if there is a planet p in this system having substantially larger uncertainty in its orbit with respect to the uncertainties in the orbits of the other planets, then the system must be treated numerically by the three-dimensional method. A `deliberate' application of the two-dimensional method in the given system, instead, reduces the accuracy of the simulation. For instance, when applying the two-dimensional method to the system HD~34445, with fixed orbit radius 6.36 for the planet g, we find $n_\mathrm{opt}=2.418$, $\Delta_\mathrm{opt} \simeq 1.2\%$, and $R_\mathrm{opt}=1.349$. So, the quantity $\Delta_\mathrm{opt}$ is $\sim 2$ times greater than the corresponding value $\simeq 0.96$ found by the three-dimensional method. Likewise, when applying the two-dimensional method to the system TRAPPIST-1, with fixed orbit radius 0.063 for the planet h, we find $n_\mathrm{opt}=2.425$, $\Delta_\mathrm{opt} \simeq 15.8\%$, and $R_\mathrm{opt}=0.116$. So, the quantity $\Delta_\mathrm{opt}$ is $\sim 2.5$ times greater than the corresponding value $\simeq 6.3$ found by the three-dimensional method (note also that the computed host star radius lies to the left of the revised interval [0.118, 0.124] discussed above).  

Concluding, we emphasize on the fact that the available observational data themselves point out the proper version of our method to be applied to a particular system. Hence, choosing a version for applying to a system seems to be a matter of astrophysical knowledge about the system.
On the other hand, if the number of the confirmed planets of a system is less than four (i.e. $N_\mathrm{P} < 4$), then the statistical procedure followed by the method for selecting the optimum model may give questionable results. It is worth clarifying here that we have not applied our method to such systems. 
  
Finally, we summarize the predicted radii (expressed in solar radii) $1.0115$ for the star Kepler-11, $1.2073$ for Kepler-90, $1.0253$ for Kepler-215, $1.0898$ for HD~10180, $1.3400$ for HD~34445, and $0.1228$ for the star TRAPPIST-1; and the predicted orbit radii (expressed in astronomical units) $6.5002$ for the planet g of the system HD~34445, and $0.0635$ for the planet h of the system TRAPPIST-1.

\begin{table}
\begin{center}
\caption{The system Kepler-11: central body $S_1$, i.e. the host star Kepler-11, and polytropic spherical shells of the planets b, c, d, e, f, and g. \label{k11}}
\begin{tabular}{lrrl} 
\hline \hline
Host star Kepler-11 -- Shell No                 & 1                         \\
\hline
            & & $R_\mathrm{q}$~~~~~ & $~~~(R_\mathrm{q})_\mathrm{u}$ \\  
\hline 
$n_\mathrm{opt}$                                & $2.833 \ \, (+00)$  \\
$\xi_{1\mathrm{opt}}$                           & $6.2983(+00)$  \\
$R_\mathrm{opt}$                                & $1.0115(+00)$  & $1.065(+00)$ &
$_{-2.2(-02)}^{+1.7(-02)}$ \\
\hline                                          
Planets of the system & \\
\hline
            & & $A$~~~~~~           & $~~~~\%D$ \\  
\hline

b -- Shell No                                 & 5                 \\
Inner radius, $\, \xi_4$                      & $8.0661(-02)$     \\              
Outer radius, $\xi_5$                         & $1.3314(-01)$     \\
Orbit radius, $\, \alpha_\mathrm{b}=\alpha_5$ & $9.1117(-02)$ & $9.10(-02)$ & $1.29(-01)$ \\
\hline

c -- Shell No                                 & 5                 \\
Orbit radius, $\, \alpha_\mathrm{c}=\alpha_\mathrm{R5}$ & $1.0680(-01)$ & $1.07(-01)$ & $1.86(-01)$ \\
\hline

d -- Shell No                                 & 6                 \\
Outer radius, $\xi_6$                         & $2.2440(-01)$     \\
Orbit radius, $\, \alpha_\mathrm{d}=\alpha_\mathrm{L6}$ & $1.4863(-01)$ & $1.55(-01)$ & $4.11(+00)$ \\   
\hline

e -- Shell No                                 & 6                 \\
Orbit radius, $\, \alpha_\mathrm{e}=\alpha_\mathrm{R6}$ & $1.9717(-01)$ & $1.95(-01)$ & $1.11(+00)$               \\ 
\hline

f -- Shell No                                    & 7                \\
Outer radius, $\xi_7$                            & $3.4952(-01)$     \\
Orbit radius, $\, \alpha_\mathrm{f}=\alpha_\mathrm{L7}$    & $2.4721(-01)$ & $2.50(-01)$ & $1.12(+00)$               \\   
\hline

g -- Shell No                                    & 8                \\
Outer radius, $\xi_8$                            & $5.0843(-01)$     \\
Orbit radius, $\, \alpha_\mathrm{g}=\alpha_\mathrm{R8}$    & $4.6604(-01)$ & $4.66(-01)$ & $9.01(-03)$               \\   
\hline

\end{tabular}
\end{center}
\end{table}

\begin{table}
\begin{center}
\caption{The system Kepler-90: central body $S_1$, i.e. the host star Kepler-90, and polytropic spherical shells of the planets b, c, i, d, e, f, g, and h. \label{k90}}
\begin{tabular}{lrrl} 
\hline \hline
Host star Kepler-90 -- Shell No                 & 1                         \\
\hline
            & & $R_\mathrm{q}$~~~~~ & $~~~~(R_\mathrm{q})_\mathrm{u}$ \\  
\hline 
$n_\mathrm{opt}$                                & $2.784 \ \, (+00)$  \\
$\xi_{1\mathrm{opt}}$                           & $6.1407(+00)$  \\
$R_\mathrm{opt}$~                               & $1.2073(+00)$  & $1.2(+00)$ &
$\pm 1.0(-01)$ \\
\hline
Planets of the system & \\
\hline
            & & $A$~~~~~~           & $~~~~\%D$ \\  
\hline

b -- Shell No                                 & 4                 \\
Inner radius, $\, \xi_3$                      & $5.2929(-02)$     \\              
Outer radius, $\xi_4$                         & $8.4588(-02)$     \\
Orbit radius, $\, \alpha_\mathrm{b}=\alpha_\mathrm{R4}$ & $7.4060(-02)$ & $7.4(-02)$ & $8.06(-02)$ \\
\hline

c -- Shell No                                 & 5                 \\
Outer radius, $\xi_5$                         & $1.4909(-01)$     \\
Orbit radius, $\, \alpha_\mathrm{c}=\alpha_\mathrm{L5}$ & $9.1432(-02)$ & $8.9(-02)$ & $2.73(+00)$ \\
\hline

i -- Shell No                                 & 6                 \\
Outer radius, $\xi_6$                         & $2.4677(-01)$     \\
Orbit radius, $\, \alpha_\mathrm{i}=\alpha_6$ & $1.9292(-01)$ & $2.028(-01)$ & $4.86(+00)$ \\
\hline

d -- Shell No                                 & 7                 \\
Outer radius, $\xi_7$                         & $3.5318(-01)$     \\
Orbit radius, $\, \alpha_\mathrm{d}=\alpha_7$ & $3.2001(-01)$ & $3.2(-01)$ & $3.22(-03)$ \\   
\hline

e -- Shell No                                 & 8                 \\
Outer radius, $\xi_8$                         & $4.4583(-01)$     \\
Orbit radius, $\, \alpha_\mathrm{e}=\alpha_\mathrm{R8}$ & $4.2058(-01)$ & $4.2(-01)$ & $1.39(-01)$               \\ 
\hline

f -- Shell No                                    & 9                \\
Outer radius, $\xi_9$                            & $6.0725(-01)$     \\
Orbit radius, $\, \alpha_\mathrm{f}=\alpha_\mathrm{L9}$    & $4.6722(-01)$ & $4.8(-01)$ & $2.66(+00)$               \\   
\hline

g -- Shell No                                    & 10                \\
Outer radius, $\xi_{10}$                            & $8.2442(-01)$     \\
Orbit radius, $\, \alpha_\mathrm{g}=\alpha_{10}$    & $7.0859(-01)$ & $7.1(-01)$ & $1.98(-01)$               \\   
\hline

h -- Shell No                                    & 11                \\
Outer radius, $\xi_{11}$                            & $1.0472(+00)$     \\
Orbit radius, $\, \alpha_\mathrm{h}=\alpha_\mathrm{R11}$    & $1.0094(+00)$ & $1.01(+00)$ & $5.90(-02)$               \\   
\hline

\end{tabular}
\end{center}
\end{table}

\begin{table}
\begin{center}
\caption{The system Kepler-215: central body $S_1$, i.e. the host star Kepler-215, and polytropic spherical shells of the planets b, c, d, and e. \label{k215}}
\begin{tabular}{lrrl} 
\hline \hline
Host star Kepler-215 -- Shell No                 & 1                         \\
\hline
            & & $R_\mathrm{q}$~~~~~ & $~~~~(R_\mathrm{q})_\mathrm{u}$ \\  
\hline 
$n_\mathrm{opt}$                                & $2.884 \ \, (+00)$  \\
$\xi_{1\mathrm{opt}}$                           & $6.4705(+00)$  \\
$R_\mathrm{opt}$                                & $1.0253(+00)$  & $1.027(+00)$ &
$\pm 2.36(-01)$ \\
\hline
Planets of the system & \\
\hline
            & & $A$~~~~~~           & $~~~~\%D$ \\  
\hline

b -- Shell No                                 & 4                 \\
Inner radius, $\, \xi_3$                      & $5.9371(-02)$     \\              
Outer radius, $\xi_4$                         & $1.0413(-01)$     \\
Orbit radius, $\, \alpha_\mathrm{b}=\alpha_\mathrm{L4}$ & $8.4007(-02)$ & $8.4(-02)$ & $7.92(-03)$ \\
\hline

c -- Shell No                                 & 5                 \\
Outer radius, $\xi_5$                         & $1.5174(-01)$     \\
Orbit radius, $\, \alpha_\mathrm{c}=\alpha_5$ & $1.1275(-01)$ & $1.13(-01)$ & $2.25(-01)$ \\
\hline

d -- Shell No                                 & 6                 \\
Outer radius, $\xi_6$                         & $2.4934(-01)$     \\
Orbit radius, $\, \alpha_\mathrm{d}=\alpha_6$ & $1.8570(-01)$ & $1.85(-01)$ & $3.78(-01)$ \\   
\hline

e -- Shell No                                 & 7                 \\
Outer radius, $\xi_7$                         & $3.9103(-01)$     \\
Orbit radius, $\, \alpha_\mathrm{e}=\alpha_7$ & $3.1097(-01)$ & $3.14(-01)$ & $9.64(-01)$               \\ 
\hline

\end{tabular}
\end{center}
\end{table}

\begin{table}
\begin{center}
\caption{The system HD~10180: central body $S_1$, i.e. the host star HD~10180, and polytropic spherical shells of the planets c, d, e, f, g, and h. \label{hd10180}}
\begin{tabular}{lrrl} 
\hline \hline
Host star HD 10180 -- Shell No                 & 1                         \\
\hline
            & & $R_\mathrm{q}$~~~~~ & $~~~~(R_\mathrm{q})_\mathrm{u}$ \\  
\hline 
$n_\mathrm{opt}$                                & $3.096 \ \, (+00)$  \\
$\xi_{1\mathrm{opt}}$                           & $7.2911(+00)$  \\
$R_\mathrm{opt}$                                & $1.0898(+00)$  & $1.109(+00)$ &
$\pm 3.6(-02)$ \\
\hline
Planets of the system & \\
\hline
            & & $A$~~~~~~           & $~~~~\%D$ \\  
\hline

c -- Shell No                                 & 3                 \\
Inner radius, $\, \xi_2$                      & $2.8497(-02)$     \\              
Outer radius, $\xi_3$                         & $8.6166(-02)$     \\
Orbit radius, $\, \alpha_\mathrm{b}=\alpha_\mathrm{R3}$ & $6.8815(-02)$ & $6.412(-02)$ & $7.32(+00)$ \\
\hline

d -- Shell No                                 & 4                 \\
Outer radius, $\xi_4$                         & $1.9484(-01)$     \\
Orbit radius, $\, \alpha_\mathrm{c}=\alpha_4$ & $1.2859(-01)$ & $1.286(-01)$ & $2.87(-03)$ \\
\hline

e -- Shell No                                 & 5                 \\
Outer radius, $\xi_5$                         & $3.7327(-01)$     \\
Orbit radius, $\, \alpha_\mathrm{d}=\alpha_5$ & $2.6900(-01)$ & $2.699(-01)$ & $3.32(-01)$ \\   
\hline

f -- Shell No                                 & 6                 \\
Outer radius, $\xi_6$                         & $6.3940(-01)$     \\
Orbit radius, $\, \alpha_\mathrm{e}=\alpha_6$ & $4.8796(-01)$ & $4.929(-01)$ & $1.00(+00)$               \\ 
\hline

g -- Shell No                                    & 8                \\
Inner radius, $\, \xi_7$                      & $1.0145(+00)$     \\              
Outer radius, $\xi_8$                         & $1.5169(+00)$     \\
Orbit radius, $\, \alpha_\mathrm{f}=\alpha_\mathrm{R9}$    & $1.3826(+00)$ & $1.427(+00)$ & $3.11(+00)$               \\   
\hline

h -- Shell No                                    & 11                \\
Inner radius, $\, \xi_{10}$                      & $2.9918(+00)$     \\
Outer radius, $\xi_{11}$                            & $4.0087(+00)$     \\
Orbit radius, $\, \alpha_\mathrm{g}=\alpha_{11}$    & $3.4616(+00)$ & $3.381(+00)$ & $2.38(+00)$               \\   
\hline

\end{tabular}
\end{center}
\end{table}

\begin{table}[h]
\begin{center}
\caption{The system HD~34445: central body $S_1$, i.e. the host star HD~34445, and polytropic spherical shells of the planets e, d, c, f, b, and g. 
\label{hd34445}}
\begin{tabular}{lrrl} 
\hline \hline
Host star HD~34445 -- Shell No                & 1                   \\
\hline
            & & $R_\mathrm{q}$~~~~~ & $~~~~(R_\mathrm{q})_\mathrm{u}$           \\  
\hline 
$n_\mathrm{opt}$                                & $2.421 \ \, (+00)$  \\
$\xi_{1\mathrm{opt}}$                           & $5.1693(+00)$       \\
$R_\mathrm{opt}$                                &$1.3400(+00)$  & $1.38(+00)$  & 
$\pm 8.0(-02)$                                                        \\
\hline
Planets of the system & \\
\hline
            & & $A$~~~~~~ & $~~~~\%D$ \\  
\hline

e -- Shell No                                 & 9                 \\
Inner radius, $\, \xi_8$                      & $2.3251(-01)$     \\              
Outer radius, $\xi_9$                         & $3.1017(-01)$     \\
Orbit radius, $\, \alpha_\mathrm{e}=\alpha_9$ & $2.6871(-01)$ & $2.687(-01)$ & $3.01(-03)$\\ 
\hline

d -- Shell No                                 & 12                \\
Inner radius, $\xi_{11}$                      & $4.5653(-01)$     \\    
Outer radius, $\xi_{12}$                      & $5.4725(-01)$     \\
Orbit radius, $\, \alpha_\mathrm{d}=\alpha_\mathrm{L12}$ & $4.8208(-01)$ & $4.817(-01)$ & $7.95(-02)$ \\
\hline

c -- Shell No                                 & 14                 \\
Inner radius, $\xi_{13}$                      & $6.3329(-01)$     \\
Outer radius, $\xi_{14}$                      & $7.4202(-01)$     \\
Orbit radius, $\, \alpha_\mathrm{c}=\alpha_\mathrm{R14}$ & $7.1559(-01)$ & $7.181(-01)$ & $3.50(-01)$ \\
\hline

f -- Shell No                                 & 20                \\
Inner radius, $\xi_{19}$                      & $1.3974(+00)$     \\
Outer radius, $\xi_{20}$                      & $1.5487(+00)$     \\
Orbit radius, $\, \alpha_\mathrm{f}=\alpha_\mathrm{R20}$ & $1.5402(+00)$ & $1.543(+00)$ & $1.84(-01)$ \\ 
\hline

b -- Shell No                                    & 24                \\
Inner radius, $\xi_{23}$                         & $2.0423(+00)$     \\
Outer radius, $\xi_{24}$                         & $2.2384(+00)$     \\
Orbit radius, $\, \alpha_\mathrm{b}=\alpha_\mathrm{L24}$ & $2.0822(+00)$ & $2.075(+00)$ & $3.45(-01)$ \\ 
\hline

Planet with substantially large   \\ 
uncertainty in its orbit radius &                                   \\
\hline
            & &$A_\mathrm{[g]q}$~~~& $~(A_\mathrm{[g]q})_\mathrm{u}$ \\  
\hline

g -- Shell No                                    & 41                \\
Inner radius, $\, \xi_{40}$                      & $6.4752(+00)$     \\
Outer radius, $\xi_{41}$                         & $6.7929(+00)$     \\
Optimum orbit radius, $\alpha_\mathrm{[g]opt}=\alpha_{L41}$ & $6.5002(+00)$ & $6.36(+00)$ & $\pm 1.02(+00)$               \\
\hline

\end{tabular}
\end{center}
\end{table}

\begin{table}[h]
\begin{center}
\caption{The system TRAPPIST-1: central body $S_1$, i.e. the host star TRAPPIST-1, and polytropic spherical shells of the planets b, c, d, e, f, g, and h. 
\label{trappist1}}
\begin{tabular}{lrrl} 
\hline \hline
Host star TRAPPIST-1 -- Shell No                & 1                   \\
\hline
            & & $R_\mathrm{q}$~~~~~ & $~~~~(R_\mathrm{q})_\mathrm{u}$           \\  
\hline 
$n_\mathrm{opt}$                                & $2.466 \ \, (+00)$  \\
$\xi_{1\mathrm{opt}}$                           & $5.2737(+00)$       \\
$R_\mathrm{opt}$                                & $1.2280(-01)$  & $1.17(-01)$  & 
$\pm 3.6(-03)$                                                        \\
\hline
Planets of the system & \\
\hline
            & & $A$~~~~~~ & $~~~~\%D$ \\  
\hline

b -- Shell No                                 & 6                 \\
Inner radius, $\, \xi_5$                      & $8.9542(-03)$     \\              
Outer radius, $\xi_6$                         & $1.2909(-02)$     \\
Orbit radius, $\, \alpha_\mathrm{b}=\alpha_\mathrm{R6}$ & $1.1154(-02)$ & $1.111(-02)$ & $3.93(-01)$\\ 
\hline

c -- Shell No                                 & 7                 \\
Outer radius, $\xi_7$                         & $1.8858(-02)$     \\
Orbit radius, $\, \alpha_\mathrm{c}=\alpha_7$ & $1.5615(-02)$ & $1.522(-02)$ & $2.66(+00)$ \\
\hline

d -- Shell No                                 & 8                 \\
Outer radius, $\xi_8$                         & $2.4562(-02)$     \\
Orbit radius, $\, \alpha_\mathrm{d}=\alpha_\mathrm{L8}$ & $2.1452(-02)$ & $2.1(-02)$ & $5.65(-02)$ \\
\hline

e -- Shell No                                 & 9                 \\
Outer radius, $\xi_9$                         & $3.0478(-02)$     \\
Orbit radius, $\, \alpha_\mathrm{e}=\alpha_\mathrm{R9}$ & $2.8111(-02)$ & $2.8(-02)$ & $1.69(-01)$ \\ 
\hline

f -- Shell No                                    & 10                \\
Outer radius, $\xi_{10}$                         & $3.8410(-02)$     \\
Orbit radius, $\, \alpha_\mathrm{f}=\alpha_\mathrm{R10}$ & $3.6699(-02)$ & $3.7(-02)$ & $1.08(+00)$ \\ 
\hline

g -- Shell No                                    & 11                \\
Outer radius, $\xi_{11}$                         & $4.5477(-02)$     \\
Orbit radius, $\, \alpha_\mathrm{g}=\alpha_\mathrm{R11}$ & $4.4225(-02)$ & $4.5(-02)$ & $1.94(+00)$               \\
\hline  
Planet with substantially large   \\ 
uncertainty in its orbit radius &                                   \\
\hline
            & &$A_\mathrm{[h]q}$~~~& $~(A_\mathrm{[h]q})_\mathrm{u}$ \\  
\hline

h -- Shell No                                    & 13                \\
Inner radius, $\, \xi_{12}$                      & $5.4921(-02)$     \\
Outer radius, $\xi_{13}$                         & $6.4397(-02)$     \\
Optimum orbit radius, $\alpha_\mathrm{[h]opt}=\alpha_{13}$ & $6.3500(-02)$ & $6.3(-02)$ & $_{-1.3(-02)}^{+2.7(-02)}$               \\
\hline

\end{tabular}
\end{center}
\end{table}

\end{document}